%% file: FiniteT_resubmit2.tex
\documentclass[twocolumn,amsmath,amssymb,aps,prl]{revtex4-1}

\usepackage{graphicx}
\usepackage{dcolumn}
\usepackage{bm}
\usepackage[colorlinks,urlcolor=blue,citecolor=blue,linkcolor=blue]{hyperref}
\usepackage{comment}
\usepackage{color}

\newcommand{\up}{\uparrow}
\newcommand{\down}{\downarrow}
\renewcommand{\k}{{\bf k}}
\newcommand{\kp}{{\bf k}^\prime}
\newcommand{\kdp}{{\bf k}^{\prime\prime}}

\newcommand{\q}{{\bf q}}

\newcommand{\0}{{\bf 0}}
\newcommand{\rmi}{{\mathrm{i}}}

\newcommand{\bra}[1]{\left\langle{#1}\right|}
\newcommand{\ket}[1]{\left|{#1}\right>}

\newcommand{\nn}{\nonumber}
\newcommand{\beq}{\begin{equation}}
\newcommand{\eeq}{\end{equation}}

\newcommand{\avbeta}[1]{\big\langle{#1}\big\rangle_\beta}

\newcommand{\bc}{\hat{\bf c}}

\DeclareMathOperator{\tr}{Tr}

\newcommand{\meera}[1]{{\color{red}#1}}

\begin{document}

\title{Variational approach for impurity dynamics at finite temperature}

\author{Weizhe Edward Liu}
\affiliation{School of Physics and Astronomy, Monash University, Victoria 3800, Australia}
\affiliation{ARC Centre of Excellence in Future Low-Energy Electronics Technologies, Monash University, Victoria 3800, Australia}
%\email{weizhe.liu@monash.edu}
\author{Jesper Levinsen}
\affiliation{School of Physics and Astronomy, Monash University, Victoria 3800, Australia}
\affiliation{ARC Centre of Excellence in Future Low-Energy Electronics Technologies, Monash University, Victoria 3800, Australia}
 %\email{jesper.levinsen@monash.edu}
\author{Meera M.~Parish}
\affiliation{School of Physics and Astronomy, Monash University, Victoria 3800, Australia}
\affiliation{ARC Centre of Excellence in Future Low-Energy Electronics Technologies, Monash University, Victoria 3800, Australia}

\date{\today}
\begin{abstract}
We present a general variational principle for the dynamics of impurity particles immersed in a quantum-mechanical medium. By working within the Heisenberg picture and constructing approximate time-dependent impurity operators, we can take the medium to be in any mixed state, such as a thermal state. Our variational method is consistent with all conservation laws and, in certain cases, it is equivalent to a finite-temperature Green's function approach. As a demonstration of our method, we consider the dynamics of heavy impurities that have suddenly been introduced into a Fermi gas at finite temperature. Using approximate time-dependent impurity operators involving only one particle-hole excitation of the Fermi sea, we find that we can successfully model the results of recent Ramsey interference experiments on $^{40}$K atoms in a $^6$Li Fermi gas [M.~Cetina et al., Science \textbf{354}, 96 (2016)]. We also show that our approximation agrees well with the exact solution for the Ramsey response of a fixed impurity at finite temperature. Our approach paves the way for the investigation of impurities with dynamical degrees of freedom in arbitrary quantum-mechanical mediums.
\end{abstract}

\maketitle

The behaviour of a quantum impurity immersed in a medium is a fundamental problem in physics, having relevance to phenomena ranging from the orthogonality catastrophe~\cite{Anderson1967} to Fermi liquid theory~\cite{baym1991landau}. In addition to displaying interesting effects in their own right, quantum impurity problems can be used to build up more complex many-body systems, such as a finite density of impurities~\cite{lobo2006}. They can also be used as a probe of correlations and entanglement in a quantum-mechanical medium~\cite{Grusdt2016}.

Recent advances in cold-atom experiments have enabled a greater variety of quantum impurity problems to be investigated. The scenario of an impurity in a Bose Einstein condensate (the so-called ``Bose polaron'') has been successfully realized in experiment~\cite{Hu2016,Jorgensen2016,Camargo2018} and could potentially provide insight into bosonic mediums in general~\cite{Tempere2009}. For instance, there is the prospect of a universal Bose polaron in the regime where the boson-impurity scattering length is tuned to infinity \cite{Yoshida2018}. Similarly, experimental investigations of impurities in a Fermi gas (i.e., ``Fermi polarons'') \cite{Schirotzek2009,Nascimbene2009,Kohstall2012,Koschorreck2012,Cetina2015doi,Cetina2016,Scazza2018,Yan2019prl} have deepened our understanding of quasiparticles in both quantum gases~\cite{Massignan_Zaccanti_Bruun,Chevy2006upd,CombescotLoboChevy2007,prokofiev2008,repulsive_polaron,Massignan2011,Schmidt2011,Vlietinck2013,Goulko2016} and the solid state~\cite{Sidler2017,Efimkin2017}. In particular, recent cold-atom experiments have observed the formation of Fermi polarons and their out-of-equilibrium dynamics~\cite{Cetina2016}, thus opening up an arena in which to explore non-equilibrium phenomena in a controlled manner.

However, a major theoretical challenge is how to include the effects of temperature when modelling the behavior of quantum impurities, since this introduces two complications. First, one must consider a medium that is in a mixed rather than a pure state. Second, one often needs to perform a thermal average over the impurity's dynamical degrees of freedom (such as the initial impurity momenta in the case of mobile impurities). Therefore, 
theoretical works on Fermi and Bose polarons at finite temperature in three dimensions have, thus far, 
%focused on 
been restricted to 
pinned impurities~\cite{Goold2011oca,Knap2012prx,Schmidt2016}, weak impurity-medium interactions~\cite{Lampo2017,Levinsen2017}, the virial expansion~\cite{CuiPRL2017}, and approximate diagrammatic approaches~\cite{Hu2017,Guenther2018,Pastukhov2018}.
Most notably, there are no exact Monte Carlo approaches for such finite-temperature %Bose and Fermi 
polarons, thus emphasizing the need for alternative methods.

In this Letter, we present a time-dependent variational principle for the quantum impurity problem that can, in principle, handle any mixed state of the medium. The key simplification is to construct approximate time-dependent impurity operators that are then applied to a static medium. We apply our variational approach to the dynamics of heavy impurities that are suddenly introduced into a Fermi gas at finite temperature. For the simplest approximation of the dynamics, our approach is equivalent to ladder diagrams within a finite-temperature Green's function approach. However, the variational calculation can be easily extended to describe more complex scenarios such as impurities that are initially entangled with the medium. Moreover, our method conserves the total probability of the system, which is not always the case in diagrammatic approximations. As a demonstration of the power and accuracy of our approach, we show that it reproduces exact results for a fixed impurity with minimal error,
and it 
%even though this scenario formally involves an infinite number of particle-hole excitations. 
%Moreover, our approach 
allows us to accurately model recent Ramsey interference experiments on $^{40}$K atoms in a $^6$Li Fermi gas~\cite{Cetina2016}, where no exact solution exists.
%Our approach compares well with exact results for a fixed impurity in a finite-temperature Fermi gas, and allows us to accurately model experiments on $^{40}$K atoms in a $^6$Li Fermi gas~\cite{Cetina2016}. %\meera{thus demonstrating the success of the }

\paragraph{Variational principle.--}
To tackle finite temperature, we separate the time dependence of the impurity dynamics from the thermal average over all states of the medium, and consider the impurity annihilation operator at time $t$ within the Heisenberg picture:
\begin{align} \label{eq:HeisPic}
\hat{c}(t) = e^{i\hat{H}t} \, \hat{c}\, \, e^{-i\hat{H}t}.
\end{align}
Here, we assume a time-independent Hamiltonian $\hat{H}$ for $t>0$, and we work in units where $\hbar$, $k_B$, and the system volume are all set to 1. For the moment, we suppress the impurity's dynamical degrees of freedom (e.g., momentum) and we assume that the impurity is non-interacting with the medium at $t=0$, i.e., $\hat{c}(0) \equiv \hat{c}$. However, the formalism can be easily generalized to include more complex initial states, as we will see below.

To proceed, we consider the Heisenberg equation of motion for the impurity operator, $ i \partial_t \hat{c}(t) = [\hat{c}(t),\hat{H}]$. The exact time-dependent operator will obey this equation; however in general we will work with operators $\hat{\mathbf{c}}(t)$ that only satisfy this approximately. Thus, inspired by other time-dependent variational principles~\cite{McLachlan1964,Basile1995}, we define an ``error'' operator $\hat{\varepsilon} (t) \equiv i \partial_t \hat{\mathbf{c}}(t) - [\hat{\mathbf{c}}(t),\hat{H}]$, and then we minimize the average quantity
\begin{align}\label{eq:error_def}
\Delta(t)=\tr\left[\hat\rho_0\, \hat{\varepsilon}(t)\,
\hat{\varepsilon}^\dag (t)\right].
\end{align}
Here, the trace is over all realizations of the medium in the absence of the impurity, %\meera{(i.e., corresponding to the impurity vacuum)}, 
and $\hat\rho_0$ is the medium density matrix.
Note that we are working in Fock space and thus the impurity operator can act directly on any particular medium state~\cite{supmat}.
Importantly, our variational approach can be applied to any mixed state of the medium, but in the following we restrict ourselves to a thermal state at temperature $T$. Then we have $\hat{\rho}_0 = \exp\big(-\beta \hat{H}_0\big) / Z_0$, with $\hat{H}_0$ the medium-only Hamiltonian, $\beta \equiv T^{-1}$, and partition function $Z_0 = \tr \big[\exp\big(-\beta \hat{H}_0\big)\big]$. In the following, we define $\langle \cdots \rangle_\beta \equiv \tr\left[\hat{\rho}_0 \cdots \right]$.

We now expand $\hat{\mathbf{c}}(t) \equiv \sum_j\alpha_{j}(t) \hat{O}_{j}$, where $\alpha_{j}(t)$ are time-dependent coefficients and $\hat{O}_{j}$ are \emph{time-independent} operators consisting of unique products of impurity and medium operators. In general there is an infinite number of such operators, and the key is to limit ourselves to states that form an appropriate variational basis for the problem at hand. Substituting the expansion into Eq.~\eqref{eq:error_def}, imposing the minimization condition $\partial \Delta/\partial\dot{\alpha}^*_{j} = 0,$ and using the orthogonality of the operators $\hat{O}_{j}$, i.e., $\langle \hat{O}_{j} \hat{O}_{l}^\dag \rangle_\beta = 0$ when $j\neq l,$ we obtain~\cite{supmat}
\begin{align} \label{eq:motionT}
i\, \dot{\alpha}_{j}(t) \, \big\langle \hat{O}_{j} \hat{O}_{j}^\dag \big\rangle_\beta
= \sum_{l} \alpha_{l}(t) \big\langle \big[ \hat{O}_{l} , \hat{H}\big] \hat{O}_{j}^\dag \big\rangle_\beta.
\end{align}
This key equation determines how the expansion coefficients $\alpha_{j}(t)$ in the approximate impurity operator $\hat{\mathbf{c}}(t)$ vary in time. As opposed to the exact Heisenberg equation of motion, we see from Eq.~\eqref{eq:motionT} that the time-dependence of the impurity operator is controlled by the mixed state of the medium. This is a natural consequence of using a truncated basis of operators within our variational approach.

From Eq.~\eqref{eq:motionT}, it is straightforward to demonstrate that our variational approach is conserving in the sense that the total probability $\avbeta{\hat{\bf c} (t) \, \hat{\bf c}^\dag (t)}$ is constant (i.e., remains 1) throughout the time evolution~\cite{supmat}. Moreover, the proof of probability conservation makes no reference to the initial conditions, and it holds even when the Hamiltonian is explicitly time dependent.

For a time-independent $\hat{H}$, we can consider the stationary solutions of Eq.~\eqref{eq:motionT}, i.e., $\alpha_{j}(t)\equiv e^{-iEt}\alpha_{j}$. In that case, we may solve a set of linear equations for the coefficients, resulting in the eigenvectors $\left\{\alpha_{j}^{(n)}\right\}$ with eigenvalues $E_n$. The associated stationary impurity operators $\hat \phi_{n}\equiv \sum_j \alpha_{j}^{(n)}\hat O_j$ may be chosen to be orthonormal under a thermal average, i.e., $\big\langle\hat \phi_{m}\hat \phi^\dag_{n}\big\rangle_\beta=\delta_{mn}$. Since the total probability is conserved, these operators provide a complete basis for the approximate impurity operators and we thus have
\begin{align}
    \hat{\mathbf{c}}(t) = \sum_n \left<
    \hat{\mathbf{c}}(0) \,\hat{\phi}_{n}^\dag\right>_\beta \hat{\phi}_{n}e^{-iE_nt},
\label{eq:opexp}
\end{align}
where the thermal average allows us to take into account the boundary condition at time $t=0$. We can then construct the relevant experimental observables by taking averages over products of the approximate impurity operators.

A particular scenario of interest is where an initially non-interacting impurity is suddenly coupled to the medium through a quench of the system parameters. The many-body response to the introduction of the impurity can be probed via Ramsey interferometry~\cite{Cetina2016}, which yields the time-dependent overlap~\cite{Goold2011oca,Knap2012prx} $S(t)\equiv \big\langle \hat{c} \, e^{i\hat H_{\mathrm{i}} t} e^{-i\hat H t} \hat{c}^\dag\big\rangle_\beta = e^{iE_{\mathrm{i}} t} \big\langle \hat{c}(t) \, \hat{c}^\dag(0)\big\rangle_\beta$, where $\hat H_{\mathrm{i}}$ is the initial non-interacting Hamiltonian and $E_{\mathrm{i}}$ is the initial energy of the impurity. This so-called Ramsey response is also intimately connected to the energy spectrum of the system, since it corresponds to the Fourier transform of the impurity spectral function~\cite{Goold2011oca,Knap2012prx}.

Finally, we emphasize that our approach may be naturally extended to systems evolving under a series of time-independent Hamiltonians (or a time-dependent Hamiltonian via Trotterization). Within each such interval, we solve for the expansion coefficients using Eq.~\eqref{eq:motionT}, and then impose the boundary conditions arising from the previous evolution via the thermal average in Eq.~\eqref{eq:opexp}.

\begin{figure*}[htb]
\centering
\includegraphics[width=\textwidth]{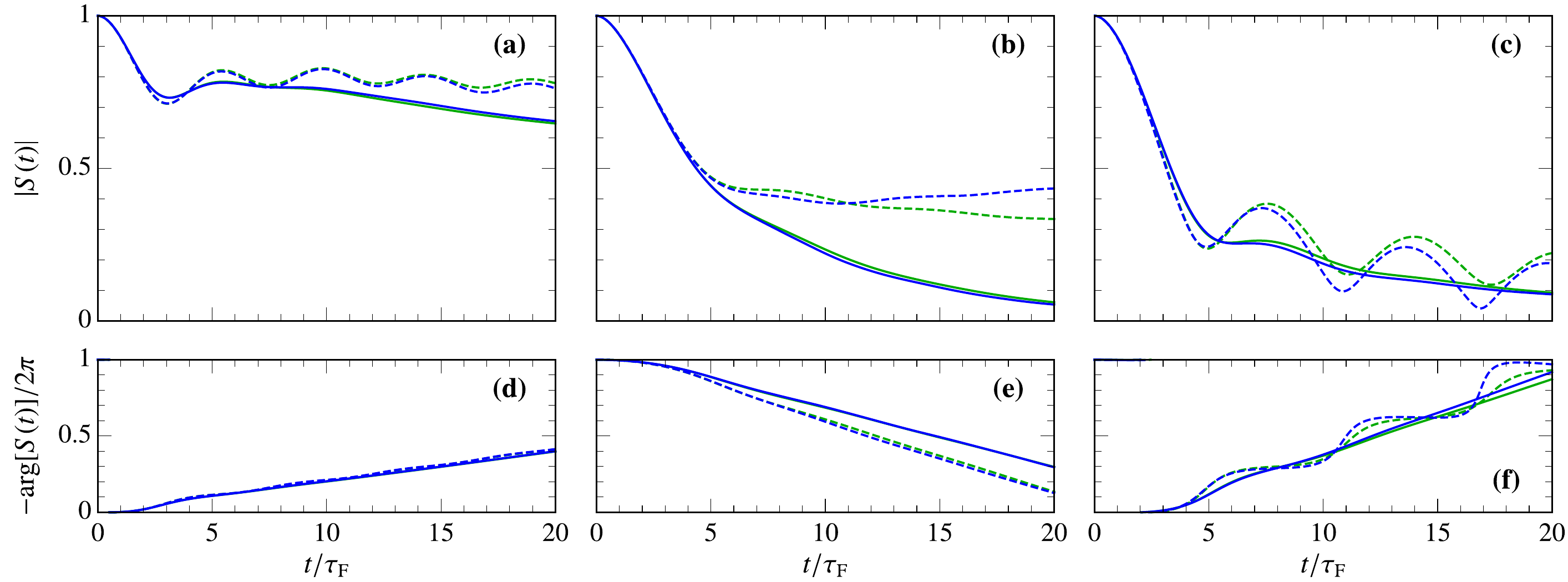}
\caption{Ramsey response of a static impurity from the variational approach (blue) and exact solution (green) for range parameter $k_{\mathrm{F}}R^* = 1$. Dashed and solid lines correspond to temperatures $T=0$ and $T=0.2 \, T_{\mathrm{F}},$ respectively. The interactions during the time evolution are (a,d) repulsive with $1/(k_{\mathrm{F}} a) = 1$; (b,e) attractive, $1/(k_{\mathrm{F}} a) = -1$; and (c,f) at unitarity $1/a = 0$.}
\label{fig:benchmarking}
\end{figure*}

\paragraph{Impurity in a Fermi sea.--}
To demonstrate the utility and accuracy of our finite-temperature variational approach, we consider the quench dynamics of a spin-$\downarrow$ impurity immersed in a spin-$\uparrow$ Fermi sea. We model the interactions using a two-channel Hamiltonian~\cite{Timmermans1999}
\begin{align}\label{eq:Hamiltonian}
\hat{H} =
&  \sum_{{\bf k}\sigma} \epsilon_{{\bf k}\sigma} \,\hat{c}^\dag_{{\bf k}\sigma}\hat{c}_{{\bf k}\sigma} + \sum_{{\bf k}} \epsilon_{{\bf k}\mathrm{M}}\,\hat{b}_{{\bf k}}^\dag \hat{b}_{{\bf k}} \nonumber \\
& + g\sum_{{\bf k}_{1}{\bf k}_{2}} \left(\hat{b}^\dag_{{\bf k}_{1}} \hat{c}_{{\bf k}_{2}\uparrow} \hat{c}_{{\bf k}_{1}-{\bf k}_{2},\downarrow} + \hat{c}^{\dag}_{{\bf k}_{1}-{\bf k}_{2},\downarrow} \hat{c}^{\dag}_{{\bf k}_{2}\uparrow}\hat{b}_{{\bf k}_{1}} \right).
\end{align}
Here, $\hat{c}^\dag_{{\bf k}\sigma}$ and $\hat{b}_{{\bf k}}^\dag$ respectively create spin-$\sigma$ fermions and closed-channel molecules with momentum ${\bf k}$, while $\epsilon_{{\bf k}\sigma} = \frac{k^{2}}{2m_{\sigma}}$ and $\epsilon_{{\bf k}\mathrm{M}} =  \frac{k^2}{2(m_\uparrow+m_\downarrow)} + \nu$, where $m_\sigma$ is the spin-$\sigma$ fermion mass and $\nu$ is the bare detuning of the closed channel. $g$ is the strength of the coupling between the open and closed channels. From the low-energy scattering amplitude at relative momentum $\k$, $f(k)=-1/(a^{-1}+R^*k^2+ik)$, the range parameter is $R^* = \frac{\pi}{m_r^2 g^2}$, while the $s$-wave scattering length $a$ is obtained via the prescription $\frac{m_r}{2\pi a} = -\frac{\nu}{g^2} + \sum_{{\bf k}}^{\Lambda} \frac{1}{\epsilon_{{\bf k}\uparrow} + \epsilon_{{\bf k}\downarrow}}$, where $m_r =  (\frac{1}{m_{\uparrow}} + \frac{1}{m_{\downarrow}})^{-1}$ is the reduced mass and $\Lambda$ is an ultraviolet cutoff that should not affect the low-energy dynamics. The relevant dimensionless quantities that parameterize the system are $1/(k_{\mathrm{F}} a)$, $k_F R^*$, $T/T_{\mathrm{F}}$ and $t/\tau_{\mathrm{F}}$, with $k_{\mathrm{F}}$ the Fermi wavenumber of the spin-$\up$ Fermi sea, while the Fermi temperature $T_{\mathrm{F}} = \frac{k_{\mathrm{F}}^2}{2m_\up}$ and the Fermi time $\tau_{\mathrm{F}} = 1/T_{\mathrm{F}}$.

To apply our variational approach, we take the approximate time-dependent operators to be of the form
\begin{align} \label{eq:op_expand}
\hat{\mathbf{c}}_{{\bf q}\downarrow} (t)  \simeq &\, \alpha_{{\bf q};0}(t) \, \hat{c}_{{\bf q}\downarrow} + \sum_{{\bf k}} \alpha_{{\bf q};{\bf k}}(t) \, \hat{c}_{{\bf k}\uparrow}^\dag \hat{b}_{{\bf q}+{\bf k}} \nonumber\\
& + \sum_{{\bf k}_1 \neq {\bf k}_2} \alpha_{{\bf q}; {\bf k}_1 {\bf k}_2}(t) \,\hat{c}_{{\bf k}_2\uparrow}^\dag \hat{c}_{{\bf k}_1\uparrow} \hat{c}_{{\bf q} - {\bf k}_1 + {\bf k}_2, \downarrow} ,
\end{align}
where ${\bf q}$ specifies the initial impurity momentum. The form of Eq.~\eqref{eq:op_expand} contains the lowest order terms one would obtain if one took $\hat{c}_{{\bf q}\downarrow}(t) = e^{i\hat{H}t}\hat{c}_{{\bf q}\downarrow}e^{-i\hat{H}t}$ and performed an expansion in $\hat{H}$. 
Additional terms in Eq.~\eqref{eq:op_expand} can similarly be obtained by considering higher order terms in the expansion,
and since the Hamiltonian preserves the particle number, all operators generated in this fashion have one and only one impurity annihilation operator (either in the open or the closed channel configuration)~\meera{\cite{supmat}}.
However, note that the approximate impurity operator in Eq.~\eqref{eq:op_expand} features time-dependent variational parameters, in contrast to the simple perturbative expansion.
This is similar in spirit to the zero-temperature variational approach to the impurity wave function first introduced in Ref.~\cite{Chevy2006upd}, and applied to impurity dynamics in Refs.~\cite{Cetina2016,Parish2016prb}. Taking the stationary condition for the operator in Eq.~\eqref{eq:op_expand} then yields the %coupled set of 
equations~\cite{supmat}:
\begin{subequations}\label{eq:kinarray}
\begin{align}
&(E-\epsilon_{{\bf q}\downarrow}) \alpha_{{\bf q};0} = g \sum_{{\bf k}} \alpha_{{\bf q};{\bf k}} \big\langle\hat c_{\k\up}^\dag c_{\k\up}\big\rangle_\beta,\\
&(E-\varepsilon_{\q;\k}) \alpha_{{\bf q};{\bf k}}
= g\alpha_{\q;0} + g \!\sum_{{\bf k}_{1}} \alpha_{{\bf q};{\bf k}_{1}{\bf k}} \big\langle\hat c_{\k_1\up} c^\dag_{\k_1\up} \big\rangle_\beta,\\
&(E-\varepsilon_{\q;\k_1 \k_2}) \, \alpha_{{\bf q};{\bf k}_{1}{\bf k}_{2}} = g\,\alpha_{{\bf q};{\bf k}_{2}},
\end{align}
\end{subequations}
where we have defined $\varepsilon_{\q;\k_1 \k_2} = \epsilon_{{\bf k}_{2}-{\bf k}_{1}+{\bf q},\downarrow} + \epsilon_{{\bf k}_{1}\uparrow} - \epsilon_{{\bf k}_{2}\uparrow}$, $\varepsilon_{\q;\k} = \epsilon_{{\bf q}+{\bf k},\mathrm{M}} - \epsilon_{{\bf k}\uparrow}$, and we will take $\big\langle\hat c_{\k\up}^\dag c_{\k\up}\big\rangle_\beta \equiv f_{\k\up}$ to be the Fermi-Dirac distribution of the $\up$ Fermi sea.

For the simplest quench scenario, where an impurity at momentum $\q$ is initially non-interacting with the Fermi sea, we have the Ramsey response 
\begin{align} \label{eq:st_q}
  S_\q(t) \! = \! e^{i\epsilon_{{\bf q}\down}t} \big\langle \hat{c}_{{\bf q}\down}(t) \hat{c}^\dag_{{\bf q}\down}\big\rangle_\beta 
  \simeq \sum_n \big|\alpha_{{\bf q};0}^{(n)} \big|^2 e^{i(\epsilon_{{\bf q}\down}- E_{{\bf q};n}) t} ,
\end{align}
where we have used Eq.~\eqref{eq:opexp} to obtain the approximate expression in terms of variational parameters. In practice, we will initially have a finite density of impurities in thermal equilibrium with the medium. Since the momentum of the impurity operator is preserved during the dynamics, we can thermally average over the initial momenta at the end of the calculation, yielding $S(t)= \left(\frac{2 \pi}{m_\down T}\right)^{3/2}\sum_{{\bf q}} e^{-\beta \epsilon_{{\bf q}\down}}S_\q(t)$. However, we stress that since we are not explicitly including correlations between impurities, the validity of such an approach is limited to a low density of impurities.

\begin{figure*}[hbt]
\centering
\includegraphics[width=0.98\textwidth]{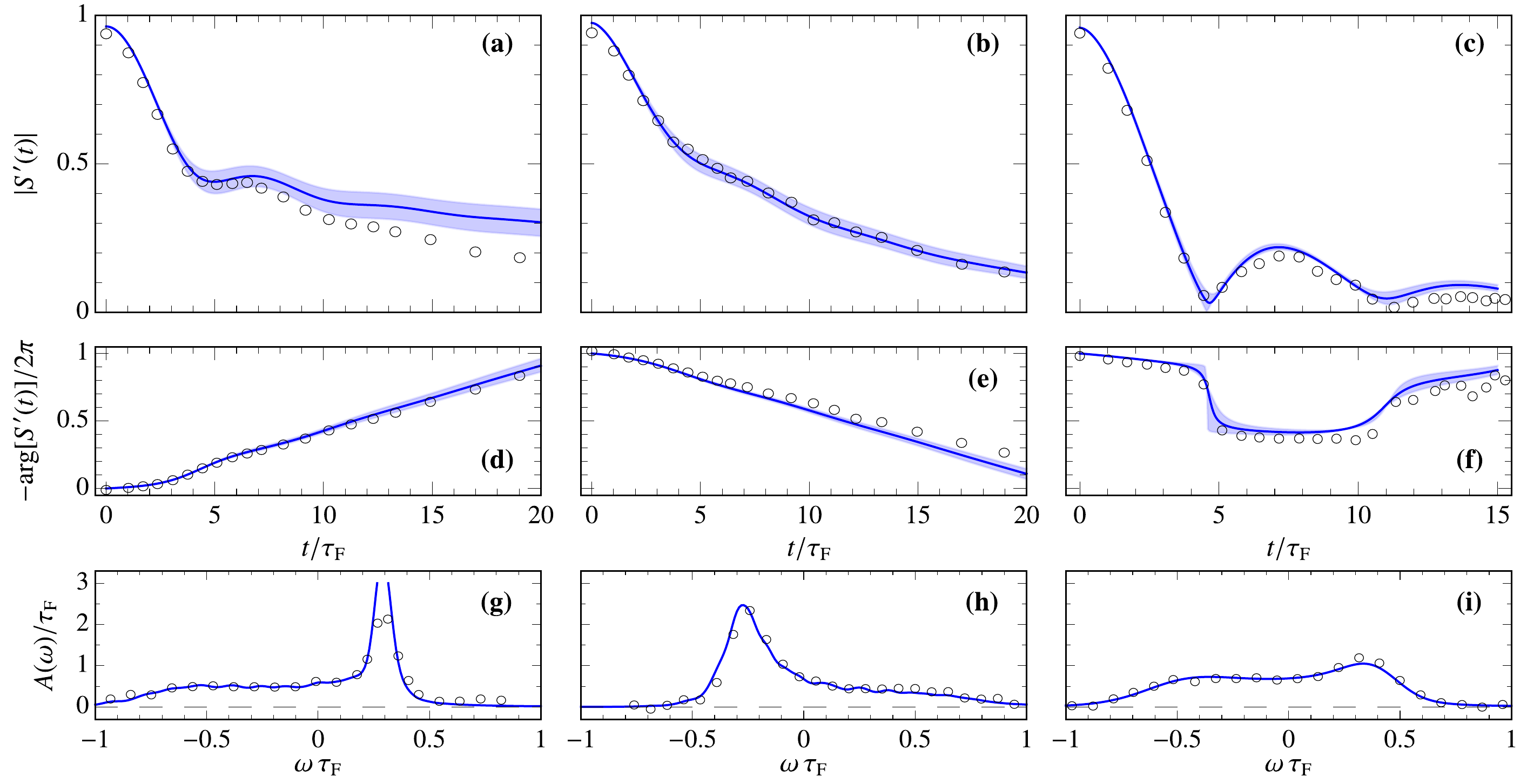}
\caption{(a-f) Ramsey response and (g-i) radio-frequency spectra of $^{40}$K impurities in a $^6$Li Fermi gas. We show the experimental results from Ref.~\cite{Cetina2016} (circles) together with results from our variational approach (solid lines), thermally averaged over impurity momentum. In (a-f) the shading corresponds to the experimental uncertainty of the scattering length~\cite{Cetina2016}. From left to right, the experimental parameters are $1/(k_{\mathrm{F}} a) = \{0.23,-0.86,-0.08\},$ $T/T_{\mathrm{F}} = \{0.17,0.16,0.18\},$ and $1/ (k_{\mathrm{F}} a_{1})= \{3.9,-5.8,-4.8\}$. In all panels $k_{\mathrm{F}}R^* = 1.1$ and $t_1 = 4.0 \, \tau_{\mathrm{F}}.$}
\label{fig:ExpFitting}
\end{figure*}

\paragraph{Dynamics of a static impurity.--}
In the particular case of a static (infinitely heavy) impurity in a Fermi sea, the quantum dynamics can be solved exactly~\cite{Levitov1996jmp,Levitov1993jetpl,Knap2012prx} since it reduces to solving for the single-particle states of the Fermi sea in the presence of a fixed potential~\cite{supmat}. 
%Therefore, this provides a benchmark for our theory. 
Since the exact solution formally involves an infinite number of excitations of the Fermi sea, this provides a highly non-trivial benchmark for our theory.
Moreover, since there is no impurity momentum, we can ignore $\q$ in Eq.~\eqref{eq:st_q} (i.e., $S(t)\equiv S_\0(t)$) and we need only consider the effects of temperature on the medium.

Figure~\ref{fig:benchmarking} displays a comparison of our variational results with the exact solution for three different interaction regimes: repulsive ($1/a>0$), attractive ($1/a<0$) and the unitary limit ($1/a = 0$). We see that our approach captures the short-time Ramsey response exactly, as expected from perturbative calculations~\cite{Parish2016prb}, and it only noticeably deviates from the exact result for times $t \gtrsim 10 \tau_{\mathrm{F}}$ at very low temperatures. Note that our single-excitation approximation cannot describe the orthogonality catastrophe~\cite{Anderson1967}, which governs the long-time behavior of the fixed impurity at $T=0$~\cite{Knap2012prx,Goold2011oca}. However, our approximation will match the exact result when $T \gg T_F$ since it contains the leading order contributions to the virial expansion. In general, thermal effects lead to an exponential decay of the amplitude $|S(t)|$ at long times~\cite{Christensen2015b,Cetina2015doi,Schmidt2018}, while the phase of $S(t)$ is determined by the dominant quasiparticle peaks in the energy spectrum, which are less sensitive to temperature. These features are all well-described within the variational approach~\cite{supmat}.
%However, we find that our approximation underestimates the decoherence rate for the repulsive case at intermediate temperatures, $0.1 \lesssim T/T_{\rm F} \lesssim 1$~\cite{supmat}.  

\paragraph{Comparison with experiment.--}
For the case of heavy $^{40}$K impurities in a $^6$Li gas~\cite{Cetina2016}, the quench dynamics involved a preparation sequence during which the impurities were weakly interacting with the Fermi gas. We model this using a two-step quench as in Ref.~\cite{Cetina2016}, which modifies the Ramsey response in Eq.~\eqref{eq:st_q} to
\begin{align}\label{eq:Sprimet_def}
S^{\prime}_{{\bf q}}(t)
= e^{i \epsilon_{{\bf q}\downarrow} (t+2t_1)} \avbeta{\hat{c}_{{\bf q}\downarrow,t_1}(t)\, \hat{c}^\dag_{{\bf q}\downarrow,-t_1}(0)},
\end{align}
where we define operators $\hat{c}_{{\bf q}\downarrow,t_1} (0) = e^{i\hat{H}_1 t_1} \hat{c}_{{\bf q}\downarrow} e^{-i\hat{H}_1 t_1}$, $\hat{c}_{{\bf q}\downarrow,t_1}(t) \equiv e^{i\hat{H} t} \hat{c}_{{\bf q}\downarrow,t_1}\, e^{-i \hat{H} t}$, and $\hat{H}_1$ is the Hamiltonian (with associated scattering length $a_1$)  applied for a time $t_1$ just before and after the time evolution governed by $\hat H$. Equation~\eqref{eq:Sprimet_def} can easily be evaluated within our variational approach by modifying the initial condition in Eq.~\eqref{eq:opexp}~\cite{supmat}. 

In Fig.~\ref{fig:ExpFitting}(a-f), we see that the calculated Ramsey response agrees remarkably well with the experimental data from Ref.~\cite{Cetina2016}, \textit{without the use of any fitting parameters}. In particular, we find that we require the thermal average over impurity momentum, as well as the thermal state of the Fermi gas, in order to accurately model the response~\cite{supmat}. Our approach also reproduces the impurity spectral function $A(\omega)=\textrm{Re}\int_{0}^\infty \frac{dt}{\pi}\,e^{i\omega t}S(t)$~\cite{supmat} measured in radio-frequency spectroscopy, as shown in  Fig.~\ref{fig:ExpFitting}(g-i). The discrepancy at long times in Fig.~\ref{fig:ExpFitting}(a) suggests that the approximation does not fully capture the decoherence rate of the repulsive branch, which is consistent with the small difference in the repulsive peak between theory and experiment in Fig.~\ref{fig:ExpFitting}(g). This intriguing result suggests that there is an additional decay channel for the repulsive branch of the heavy impurity that is not present in the infinite-mass case~\cite{supmat}.
%Fig.~\ref{fig:benchmarking}. Moreover, for the calculated spectra, we find that we must apply an additional broadening to the repulsive peak to achieve the excellent theory-experiment agreement in Fig.~\ref{fig:ExpFitting}(g)~\cite{supmat}.

\paragraph{Relationship to diagrammatic approaches.--}
Solving Eqs.~\eqref{eq:kinarray} for the energy yields the expression
\begin{align} \notag
    E = \epsilon_{{\bf q}\downarrow} + \sum_{{\bf k}_2} f_{\k_2\up} \left[\frac{E-\varepsilon_{\q;\k_2}}{g^2} - \sum_{\k_1} \frac{1-f_{\k_1\up}}{E-\varepsilon_{\q;\k_1 \k_2}} \right]^{-1} ,
\end{align}
which corresponds to the pole of the impurity Green's function, $E = \epsilon_{{\bf q}\downarrow} + \Sigma(\q,E)$, where $\Sigma(\q,E)$ is the impurity self energy calculated using ladder diagrams at finite temperature~\cite{Hu2017}. Therefore, our variational method is equivalent to a finite-temperature Green's function approach --- indeed, the Ramsey response in Eq.~\eqref{eq:st_q} is simply proportional to the time-dependent impurity Green's function. However, our formulation has the advantage that it can be readily adapted to describe more complex dynamics, such as the two-step quench in Eq.~\eqref{eq:Sprimet_def} or Rabi oscillations~\cite{Parish2016prb}.

\paragraph{Conclusions.--}
We have developed a general variational approach for impurity dynamics and we have used it to successfully model a heavy impurity in a Fermi gas at finite temperature. Our results suggest that the dynamics observed in experiment is well described by approximations with a single excitation of the Fermi sea, and that the effect of the impurity mass is masked by residual interactions during the preparation, as well as by thermal fluctuations. Our method paves the way for further investigations of quantum impurities, involving different variational operators (e.g., derived from coherent states) or other scenarios, such as the Bose polaron.

%%%%%%%%%%%%%%%%%%%%%%%%%%%%%%%%%%%%%%%%%%%%%%%%%
\acknowledgments
We are grateful to G.~Bruun for fruitful discussions, and we thank M.~Cetina for providing us with the experimental data. JL is supported through the Australian Research Council Future Fellowship FT160100244. We acknowledge support from the Australian Research Council Centre of Excellence in Future Low-Energy Electronics Technologies (CE170100039).

%%%%%%%%%%%%%%%%%%%%%%%%%%%%%%%%%%%%%%%%%%%%%%%%%
%\bibliography{references}
\input{FiniteT_resubmit2.bbl}

%%%%%%%%
%%%%%%%%
%%%%%%%%
%%%%%%%%

\clearpage % Start of the Supplementary Material

\renewcommand{\theequation}{S\arabic{equation}}
\renewcommand{\thefigure}{S\arabic{figure}}

\onecolumngrid

\newpage

\setcounter{equation}{0}
\setcounter{figure}{0}
\setcounter{page}{1}

\section*{SUPPLEMENTAL MATERIAL: ``Variational approach for impurity dynamics at finite temperature''}

\begin{center}
Weizhe Edward Liu, Jesper Levinsen, and Meera M.~Parish,\\
\begin{small}
\textit{School of Physics and Astronomy, Monash University, Victoria 3800, Australia
and\\ 
ARC Centre of Excellence in Future Low-Energy Electronics Technologies, Monash University, Victoria 3800, Australia}
\end{small}
\end{center}

\subsection{Error operator and error quantity}

In the main text, we define the error operator 
\begin{align}
\hat{\varepsilon} (t) \equiv i \partial_t \hat{\mathbf{c}}(t) - [\hat{\mathbf{c}}(t),\hat{H}],
\label{eq:errorop}
\end{align}
as the error in the Heisenberg equation of motion for the impurity operator within our variational approach. We then minimize the error quantity
\begin{align}\label{eq:error_defSM}
\Delta(t)=\tr\left[\hat\rho_0\, \hat{\varepsilon}(t)\,
\hat{\varepsilon}^\dag (t)\right],
\end{align}
with respect to the time derivative of the variational parameters in the impurity operator,
\begin{align}
    \hat{\mathbf{c}}(t) \equiv \sum_j\alpha_{j}(t) \hat{O}_{j}.
    \label{eq:impexp}
\end{align}
As discussed in the main text, the trace in Eq.~\eqref{eq:error_defSM} should be taken over all realizations of the medium in the absence of the impurity. We can thus choose a complete set of medium states, $\{\ket{{\rm med};0} \}$, which are eigenstates of the medium-only Hamiltonian $\hat H_0$, with $\hat H_0\ket{{\rm med};0}=E_{\rm med}\ket{{\rm med};0}$, and which are vacuum states with respect to the impurity degrees of freedom. Then the error quantity $\Delta$ is explicitly a sum over all medium configurations:
\begin{align}
\Delta(t)=\frac1{Z_0}\sum_{\{\rm med\}}\bra{{\rm med};0}e^{-\beta E_{\rm med}}
\hat{\varepsilon}(t)\,
\hat{\varepsilon}^\dag (t)\ket{{\rm med};0}.
\label{eq:Deltaexplicit}
\end{align}
Since we are working in Fock space, the impurity operator can act directly on the medium states; for instance we have $c^\dag_\q\ket{{\rm med};0}=\ket{{\rm med};\q}$ and $c_\q\ket{{\rm med};0}=0$. Likewise, the error operator in Eq.~\eqref{eq:errorop} consists of impurity and medium operators that act directly on $\ket{{\rm med};0}$. 
However, it is also important to note that we only consider variational operators that are number conserving and thus the operators $\hat{O}_{j}$ in Eq.~\eqref{eq:impexp} all have just one impurity annihilation operator. This guarantees that $\hat{O}_{j}$ does not annihilate the impurity vacuum in Eq.~\eqref{eq:error_defSM} and thus $\Delta(t)$ is not trivially zero for $t>0$.

Specifically, for the model in Eq.~\eqref{eq:Hamiltonian} of the main text, the error operator in Eq.~\eqref{eq:errorop} is a superposition of impurity annihilation operators in either the open or closed channel. Thus we have
\begin{align}
    \hat \varepsilon(t)=\sum_\k \hat M_\k(t)\hat c_{\k\down}+\sum_\k \hat N_\k(t)\hat b_\k,
\end{align}
where $\hat M_\k(t)$ and $\hat N_\k(t)$ are operators that act only on the medium.
Therefore, the error quantity is the real, non-negative function
\begin{align}
    \Delta(t)=\sum_\k \tr \left[\rho_0 \hat M_\k(t)\hat M_\k^\dag(t)\right]+
    \sum_\k \tr \left[\rho_0 \hat N_\k(t)\hat N_\k^\dag(t)\right]\geq0,
\end{align}
and $\Delta(t)=0$ if and only if each operator $\hat M_\k(t)$ and $\hat N_\k(t)$ is individually zero, i.e., if and only if the error operator is identically zero.

\subsection{Variational principle at nonzero temperature}

Here we discuss how to derive Eq.~\eqref{eq:motionT} from the main text. Substituting the operator expansion $\bc (t) =\sum_j \alpha_j (t) \hat{O}_j$ into Eq.~(\ref{eq:error_def}), we have
\begin{align}
\Delta (t) = & \, \big\langle \hat{\varepsilon} (t)\, \hat{\varepsilon}^\dag (t) \big\rangle_\beta
=  \sum_{jl} \big\langle \big\{ i \dot{\alpha}_{j} (t) \hat{O}_{j} - \alpha_{j} (t) [\hat{O}_{j},\hat{H}]\big\} \big\{ - i \dot{\alpha}_{l}^* (t) \hat{O}_{l}^\dag + \alpha_{l}^* (t) [\hat{O}_{l}^\dag,\hat{H}] \big\}\big\rangle_\beta. 
\end{align}
The condition $\partial \Delta (t) /\partial\dot{\alpha}^*_{l} (t) = 0$ then yields
\begin{align}
\sum_{j} \Big[\dot{\alpha}_{j} (t) \big\langle \hat{O}_{j} \hat{O}_{l}^\dag \big\rangle_\beta + i \, \alpha_{j} (t) \big\langle \big[\hat{O}_{j}, \hat{H} \big] \hat{O}_{l}^\dag \big\rangle_\beta \Big] = 0.
\end{align}
Since each operator $\hat{O}_{j}$ consists of a unique product of operators, we have $\big\langle \hat{O}_{j} \hat{O}_{l}^\dag\big\rangle_\beta = 0$ if $j\neq l.$ Thus,
\begin{align}
i \,\dot{\alpha}_{l} (t) \big\langle \hat{O}_{l} \hat{O}_{l}^\dag \big\rangle_\beta = \sum_{j} \alpha_{j} (t) \big\langle \big[\hat{O}_{j}, \hat{H} \big] \hat{O}_{l}^\dag \big\rangle_\beta, 
\label{eq:alphaeq}
\end{align}
which, after interchanging labels, is Eq.~\eqref{eq:motionT} in the main text.

\subsection{Conservation of Probability}

To demonstrate that the total probability $\avbeta{\hat{\bf c} (t) \, \hat{\bf c}^\dag (t)}$ is constant, it is sufficient to show that its time derivative vanishes. According to the operator expansion in Eq.~\eqref{eq:impexp}, the time derivative becomes
\begin{align}
 \frac{\mathrm{d} \big\langle \bc (t) \, \bc^\dag (t)\big\rangle_\beta}{\mathrm{d} t}
= &\sum_{j} \big\langle \hat{O}_{j} \hat{O}_{j}^\dag \big\rangle_\beta \left(\dot{\alpha}_{j} (t) \alpha_{j}^* (t) + \alpha_{j} (t) \dot{\alpha}_{j}^* (t) \right) \nn \\ 
= & - i \sum_{jl}\left[ - \alpha_{l} (t) \alpha_{j}^* (t) \big\langle \hat{H} \hat{O}_{l} \hat{O}_{j}^\dag \big\rangle_\beta + \alpha_{l} (t) \alpha_{j}^* (t) \big\langle \hat{O}_{l} \hat{O}_{j}^\dag \hat{H} \big\rangle_{\beta}\right] 
%\nonumber\\
%=&\, 0,\nn
\end{align}
where we have used the orthogonality of the operators $\hat{O}_{j}$ in the first step, and Eq.~\eqref{eq:alphaeq} in the second step. We have also cancelled terms containing $\big\langle \hat O_l \hat H\hat O_j^\dag\big\rangle_\beta$.

To show that $\mathrm{d} \big\langle \bc (t) \, \bc^\dag (t)\big\rangle_\beta/\mathrm{d} t=0$ we thus need $\big\langle \hat{H} \hat{O}_{l} \hat{O}_{j}^\dag \big\rangle_\beta-\big\langle  \hat{O}_{l} \hat{O}_{j}^\dag \hat{H} \big\rangle_\beta=0$. To see that this is indeed the case, note that $\hat H$ reduces to $\hat H_0$ when acting on the impurity vacuum, and that $\rho_0$ acting on the impurity vacuum returns an impurity vacuum state [see, e.g., Eq.~\eqref{eq:Deltaexplicit}]. Therefore, we have $\big\langle \hat{H} \hat{O}_{l} \hat{O}_{j}^\dag \big\rangle_\beta-\big\langle  \hat{O}_{l} \hat{O}_{j}^\dag \hat{H} \big\rangle_\beta=\big\langle \hat{H_0} \hat{O}_{l} \hat{O}_{j}^\dag \big\rangle_\beta-\big\langle  \hat{O}_{l} \hat{O}_{j}^\dag \hat{H_0} \big\rangle_\beta$ which is 0 since the medium-only Hamiltonian $H_0$ and $\rho_0$ share a complete set of (impurity vacuum) eigenstates, i.e., the medium states $\{\ket{\rm{med};0}\}$. Explicitly, using the construction in %in analogy to 
Eq.~\eqref{eq:Deltaexplicit}, we have
\begin{align}
    \big\langle \hat{H} \hat{O}_{l} \hat{O}_{j}^\dag \big\rangle_\beta-\big\langle  \hat{O}_{l} \hat{O}_{j}^\dag \hat{H} \big\rangle_\beta & 
    =\frac1{Z_0}\sum_{\{\rm med\}}\bra{{\rm med};0}e^{-\beta E_{\rm med}}\left(\hat{H} \hat{O}_{l} \hat{O}_{j}^\dag - \hat{O}_{l} \hat{O}_{j}^\dag \hat{H}\right)\ket{{\rm med};0}\nn \\
    & = \frac1{Z_0}\sum_{\{\rm med\}}\bra{{\rm med};0}e^{-\beta E_{\rm med}}\left(\hat{H_0} \hat{O}_{l} \hat{O}_{j}^\dag - \hat{O}_{l} \hat{O}_{j}^\dag \hat{H_0}\right)\ket{{\rm med};0} \nn \\
    & = \frac1{Z_0}\sum_{\{\rm med\}}\bra{{\rm med};0}e^{-\beta E_{\rm med}}\left(E_{\rm med} \hat{O}_{l} \hat{O}_{j}^\dag -\hat{O}_{l} \hat{O}_{j}^\dag E_{\rm med}\right)\ket{{\rm med};0}\nn \\ & = 0.
\end{align}
Thus, the total probability $\avbeta{\hat{\bf c} (t) \, \hat{\bf c}^\dag (t)}$ is conserved within our variational approach.

%In the last step, 
%\meera{we have used the fact that $\hat H$ reduces to $\hat H_0$ when acting on the impurity vacuum, and that $\hat H_0$ in turn commutes with $\rho_0$. That is, we have $\big\langle \hat{O}_{l} \hat{O}_{j}^\dag \hat{H} \big\rangle_{\beta}=\big\langle \hat{O}_{l} \hat{O}_{j}^\dag \hat{H}_0 \big\rangle_{\beta}=\big\langle \hat{H}_0  \hat{O}_{l} \hat{O}_{j}^\dag \big\rangle_{\beta}$. Likewise, since $\rho_0$ acting on an impurity vacuum state $\ket{\rm{med};0}$ yields another impurity vacuum state, we have $\big\langle \hat{H} \hat{O}_{l} \hat{O}_{j}^\dag \big\rangle_{\beta}=\big\langle \hat{H}_0 \hat{O}_{l} \hat{O}_{j}^\dag \big\rangle_{\beta}$.}

%we use the cyclic property of the trace and the fact that $\hat H$ commutes with $\rho_0$.

\subsection{Variational equations for the Fermi polaron}

We now specialize to a fermionic medium, and derive the variational equations for impurity operators of the form in Eq.~\eqref{eq:op_expand}. In this part, we will omit the hat ``$\hat{\,\,\,\,}$'' for operator $O$, $H$, $c$ and $b.$ We also omit the $\q$ in the subscripts of $\alpha$ and $O.$ From Eq.~\eqref{eq:motionT}, we have
\begin{align}
E \, \alpha_j \avbeta{O_j O_j^\dag} = & \sum_l \alpha_l \avbeta{[O_l,H_0] O_j^\dag} + \sum_l \alpha_l \avbeta{O_l (H-H_0) O_j^\dag}.
\label{eq:sm:motionT}
\end{align}
Note that the form of this equation ensures that we measure the polaron energy with respect to that of the unperturbed Fermi sea. In the case of the Fermi polaron, the indices on the generalized operators correspond to $j,l=\{0,\k,\k_1 \k_2\}$. Here we list the expressions for all variables in the above equation:
\beq\label{eq:sm:allvariables}
\left\{\begin{array}{l}
O_0 = c_{\q\down};\quad O_\k = c_{\k \up}^\dag b_{\q + \k}; \quad O_{\k_1 \k_2} = c_{\k_2 \up}^\dag c_{\k_1\up} c_{\q - \k_1 + \k_2,\down}, \\[2ex]
\displaystyle H - H_0 = \sum_{\k} \epsilon_{\k\down} c_{\k\down}^\dag c_{\k\down} + \sum_{\k} \epsilon_{\k\mathrm{M}} b_{\k}^\dag b_{\k} + g \sum_{\k_1\k_2} \big( b_{\k_1}^\dag c_{\k_2\up}c_{\k_1 - \k_2,\down} + c_{\k_1 - \k_2,\down}^\dag c_{\k_2\up}^\dag b_{\k_1} \big), 
\end{array}\right.
\eeq
where $\k_1\neq \k_2$ in $O_{\k_1 \k_2}$ and $H_0 = \sum_{\k} \epsilon_{\k\up} c_{\k \up}^\dag c_{\k \up}.$ Furthermore, we have $\avbeta{c_{\k \up}^\dag c_{\k\up}} = f_{\k \up},$ $\avbeta{c_{\q\down}c_{\q\down}^\dag} = 1,$ and $\avbeta{b_\q b_\q^\dag} = 1,$ where $f_{\k \up}$ is the Fermi-Dirac distribution as mentioned in the main text. 

Note that we can use Wick's theorem~\cite{Fetter} to rewrite thermal averages over several operators in terms of products of averages over operator pairs. For instance, $\avbeta{c_{\q\down} b_{\kp}^\dag b_{\kp} c_{\q\down}^\dag} = \avbeta{b_{\kp}^\dag b_{\kp}} \avbeta{c_{\q\down} c_{\q\down}^\dag}=0.$ As a non-trivial example, in the following we encounter thermal averages such as $\avbeta{c_{\k_1\up}^\dag c_{\k_2\up}c^\dag_{\k_3\up}c_{\k_4\up}}$ where $\k_1\neq \k_2$ and $\k_3\neq \k_4$. We thus have
\begin{align}
    \avbeta{c_{\k_1\up}^\dag c_{\k_2\up}c^\dag_{\k_3\up}c_{\k_4\up}} & = \avbeta{c_{\k_1\up}^\dag c_{\k_1\up}c_{\k_2\up} c_{\k_2\up}^\dag}\delta_{\k_1\k_4}\delta_{\k_2\k_3}.
\end{align}
We then use
\begin{align}
    \avbeta{c^\dag_{\k_1\up}c_{\k_1\up}c_{\k_2\up}c^\dag_{\k_2\up}} &= \tr[{e^{-\beta H_0}c^\dag_{\k_1\up}c_{\k_1\up} c_{\k_2\up}c_{\k_2\up}^\dag}] \nn \\[1ex]
    &=\tr[{c_{\k_2\up}^\dag e^{-\beta H_0}c^\dag_{\k_1\up}c_{\k_1\up}c_{\k_2\up}}] \nn \\[1ex]
    &=e^{\beta \epsilon_{\k_2\up}}\tr[{e^{-\beta H_0}c_{\k_2\up}^\dag c^\dag_{\k_1\up}c_{\k_1\up}c_{\k_2\up}}] \nn \\[1ex]
    &=e^{\beta \epsilon_{\k_2\up}}\avbeta{c^\dag_{\k_1\up}c_{\k_1\up}}-e^{\beta \epsilon_{\k_2\up}}\avbeta{c^\dag_{\k_1\up}c_{\k_1\up}c_{\k_2\up}c_{\k_2\up}^\dag} \nn \\[1ex]
    &=\avbeta{c^\dag_{\k_1\up}c_{\k_1\up}}\avbeta{c_{\k_2\up}c^\dag_{\k_2\up}},
\end{align}
where in the last step we have rearranged the equation and used $\avbeta{c_{\k \up}^\dag c_{\k\up}} = f_{\k \up}$.

To derive the variational equations, we then consider these term by term:
\begin{itemize}
    \item $j=0$ in Eq.~\eqref{eq:sm:motionT}. Writing out Eq.~\eqref{eq:sm:motionT} with the help of Eq.~\eqref{eq:sm:allvariables}, we have
\end{itemize}
\begin{align}
E \, \alpha_0 = \alpha_0 \avbeta{c_{\q\down} \sum_{\kp} \epsilon_{\kp\down} c_{\kp\down}^\dag c_{\kp\down} c_{\q\down}^\dag} + \sum_{\k} \alpha_\k \avbeta{c_{\k\up}^\dag b_{\q+\k} \,g \sum_{\kp_1\kp_2} b_{\kp_1}^\dag c_{\kp_2\up}c_{\kp_1 - \kp_2,\down} c_{\q\down}^\dag} = \epsilon_{\q\down}\alpha_0 + g \sum_\k \alpha_{\k} f_{\k\up}. \nonumber
\end{align}
\begin{itemize}
\item $j = \k:$
\end{itemize}
\begin{align}
E \, \alpha_{\k} \avbeta{c_{\k\up}^\dag c_{\k\up}} = & \sum_l \alpha_l \avbeta{[O_l,\sum_{\kp} \epsilon_{\kp\up} c_{\kp \up}^\dag c_{\kp \up}] b_{\q +\k}^\dag c_{\k\up}} \label{eq:sm:alpha10}\\
& + \sum_l \alpha_l \avbeta{O_l \Big(\sum_{\kp} \epsilon_{\kp\down} c_{\kp\down}^\dag c_{\kp\down} + \sum_{\kp} \epsilon_{\kp\mathrm{M}} b_{\kp}^\dag b_{\kp} + g \sum_{\kp_1\kp_2} \, c_{\kp_1 - \kp_2,\down}^\dag c_{\kp_2\up}^\dag b_{\kp_1} \Big) b_{\q +\k}^\dag c_{\k\up}}.
\label{eq:sm:alpha11}
\end{align}
In Eq.~\eqref{eq:sm:alpha10} only terms including $O_{\k}$ survive from the thermal average, which gives $- \alpha_\k \epsilon_{\k\up} \avbeta{c_{\k\up}^\dag c_{\k\up}}.$ In Eq.~\eqref{eq:sm:alpha11}, all $O_l$ terms survive from the thermal average but they pick out different terms from the Hamiltonian:
\begin{align}
\text{\eqref{eq:sm:alpha11}}
= & \, \alpha_0 \avbeta{c_{\q\down}\, g \sum_{\kp_1\kp_2} c_{\kp_1 - \kp_2,\down}^\dag c_{\kp_2\up}^\dag b_{\kp_1} b_{\q +\k}^\dag c_{\k\up}} + \sum_{\kdp} \alpha_{\kdp}\avbeta{c_{\kdp\up}^\dag b_{\q+\kdp}  \sum_{\kp} \epsilon_{\kp\mathrm{d}} b_{\kp}^\dag b_{\kp} b_{\q +\k}^\dag c_{\k\up}} \nonumber \\
&+ \sum_{\k_1\k_2} \alpha_{\k_1\k_2} \avbeta{c_{\k_2 \up}^\dag c_{\k_1\up} c_{\q - \k_1 + \k_2,\down} \, g \sum_{\kp_1\kp_2} c_{\kp_1 - \kp_2,\down}^\dag c_{\kp_2\up}^\dag b_{\kp_1} b_{\q +\k}^\dag c_{\k\up}} \nonumber \\
= & \, g \,\alpha_0\avbeta{c_{\k\up}^\dag c_{\k\up}} + \epsilon_{\q+\k,\mathrm{M}}\alpha_\k \avbeta{c_{\k\up}^\dag c_{\k\up}} + g\sum_{\k_1} \alpha_{\k_1\k} \avbeta{c_{\k\up}^\dag c_{\k\up}} \avbeta{c_{\k_1\up} c_{\k_1 \up}^\dag}. \nonumber
\end{align}
%
%\clearpage

\begin{itemize}
\item $j = \k_1 \k_2:$
\end{itemize}
\begin{align}
E\, \alpha_{\k_1\k_2} \avbeta{c_{\k_1\up} c_{\k_1\up}^\dag} \avbeta{c_{\k_2\up}^\dag c_{\k_2\up}} = & \sum_l \alpha_l \avbeta{[O_l,\sum_{\kp} \epsilon_{\kp\up} c_{\kp \up}^\dag c_{\kp \up}] c_{\q - \k_1 + \k_2,\down}^\dag c_{\k_1 \up}^\dag c_{\k_2 \up}} \label{eq:sm:alpha20} \\
& + \!\sum_l \!\alpha_l \avbeta{O_l\Big[\!\sum_{\kp} \!\epsilon_{\kp\down} c_{\kp\down}^\dag c_{\kp\down} \! + g \! \sum_{\kp_1\kp_2} \!\! b_{\kp_1}^\dag c_{\kp_2\up}c_{\kp_1 - \kp_2,\down}\!\Big] c_{\q - \k_1 + \k_2,\down}^\dag c_{\k_1 \up}^\dag c_{\k_2 \up}}\label{eq:sm:alpha21}
\end{align}
It is straightforward to see that Eq.~\eqref{eq:sm:alpha20} gives $\alpha_{\k_1\k_2}\big(\epsilon_{\k_1\up} -\epsilon_{\k_2\up}\big) \avbeta{c_{\k_1\up} c_{\k_1\up}^\dag}  \avbeta{c_{\k_2\up}^\dag c_{\k_2\up}}$, while in Eq.~\eqref{eq:sm:alpha21} the $O_\k$ and $O_{\k_1\k_2}$ terms will survive:
\begin{align}
\text{\eqref{eq:sm:alpha21}} = & \sum_{\k} \alpha_{\k} \avbeta{c_{\k\up}^\dag b_{\q+\k} \, g \! \sum_{\kp_1\kp_2} b_{\kp_1}^\dag c_{\kp_2\up}c_{\kp_1 - \kp_2,\down}  c_{\q - \k_1 + \k_2,\down}^\dag c_{\k_1 \up}^\dag c_{\k_2 \up}} \nonumber \\
& + \sum_{\kp_1\kp_2}\alpha_{\kp_1\kp_2} \avbeta{c_{\kp_2 \up}^\dag c_{\kp_1\up} c_{\q - \kp_1 + \kp_2,\down} \sum_{\kp} \!\epsilon_{\kp\down} c_{\kp\down}^\dag c_{\kp\down} c_{\q - \k_1 + \k_2,\down}^\dag c_{\k_1 \up}^\dag c_{\k_2 \up}} \nonumber \\
= & \, g \, \alpha_{\k_2} \avbeta{c_{\k_1\up} c_{\k_1\up}^\dag} \avbeta{c_{\k_2\up}^\dag c_{\k_2\up}} + \epsilon_{\q-\k_1+\k_2,\down} \avbeta{c_{\k_1\up} c_{\k_1 \up}^\dag}\avbeta{c_{\k_2 \up}^\dag c_{\k_2 \up}}. \nonumber
\end{align}

In summary, we put $\q$ back into the subscript of $\alpha_{j}$ and we have
\beq\label{eq:sm:kinarray}\left\{
\begin{array}{l}
\displaystyle E \, \alpha_{{\bf q};0} = \epsilon_{{\bf q}\downarrow} \alpha_{{\bf q};0} + g \sum_{{\bf k}} \alpha_{{\bf q};{\bf k}} f_{{\bf k}\uparrow},\\[2ex]
\displaystyle E \, \alpha_{{\bf q};{\bf k}} = \left(\epsilon_{{\bf q}+{\bf k},\mathrm{M}} - \epsilon_{{\bf k}\uparrow} \right) \alpha_{{\bf q};{\bf k}} + g\,\alpha_{\q;0} + g \sum_{{\bf k}_{1}} \alpha_{{\bf q};{\bf k}_{1}{\bf k}} \avbeta{c_{\k_1\up} c_{\k_1 \up}^\dag},\\[2ex]
E \, \alpha_{{\bf q};{\bf k}_{1}{\bf k}_{2}} = (\epsilon_{{\bf k}_{2}-{\bf k}_{1}+{\bf q},\downarrow} + \epsilon_{{\bf k}_{1}\uparrow} - \epsilon_{{\bf k}_{2}\uparrow}) \alpha_{{\bf q};{\bf k}_{1}{\bf k}_{2}} + g\,\alpha_{{\bf q};{\bf k}_{2}}.
\end{array}\right.\eeq
where we have divided out all the common factors arising from the thermal averages. This recovers Eq.~\eqref{eq:kinarray} of the main text. Furthermore, we note that the orthonormality of the stationary operators, $\avbeta{\hat{\phi}_{m} \hat{\phi}_{n}^\dag} = \delta_{m n}$, means that
\beq
\alpha_{{\bf q};0}^{(m)} \alpha_{{\bf q};0}^{(n)\, *} + \sum_{{\bf k}} \alpha^{(m)}_{{\bf q};{\bf k}} \alpha^{(n)\,*}_{{\bf q};{\bf k}} f_{{\bf k}} + \sum_{{\bf k}_{1}{\bf k}_{2}} \alpha^{(m)}_{{\bf q};{\bf k}_{1}{\bf k}_{2}} \alpha^{(n)\, *}_{{\bf q};{\bf k}_{1}{\bf k}_{2}} \left(1-f_{{\bf k}_{1}} \right) f_{{\bf k}_{2}} =\delta_{m n}.
\eeq

\subsection{Exact solution for a static impurity}

In the case of a static impurity, the momenta of impurities and closed-channel molecules become trivial and then the explicit expressions of $\hat{H}_\rmi$ and $\hat{H}$ may be written as
\begin{align}
\hat{H}_\rmi & = \nu \, \hat{b}^\dag \hat{b} + \sum_\k \epsilon_{\k\up} \hat{c}_{\k\up}^\dag \hat{c}_{\k\up} 
= \nu \, \hat{b}^\dag \hat{b} + \sum_{lm} \int_0^\infty \frac{{\rm d} k}{2\pi} \frac{k^2}{2m} \hat{c}_{klm} \hat{c}_{klm}^\dag, \nonumber \\
\hat{H} & = \hat{H}_\rmi  + g \sum_\k (\hat{b}^\dag \hat{c}_{\k \up} + \hat{c}_{\k\up}^\dag \hat{b}) = \hat{H}_\rmi + g \int_0^\infty \frac{k \, {\rm d} k}{2 \pi\sqrt{\pi}} (\hat{b}^\dag \hat{c}_{k00} + \hat{c}_{k00}^\dag \,\hat{b}), \label{eq:sm:FDAimp}
\end{align}
where we expand $\hat{c}_{\k\up} = \sum_{lm} (2\pi/k) Y_{lm} (\theta,\phi) \hat{c}_{klm}$ with the help of the spherical harmonics $Y_{lm} $ whose orthonormality condition is $\int \int Y_{lm} (\theta,\phi) Y_{l^\prime m^\prime} (\theta,\phi) \sin\theta \, {\rm d} \theta \,{\rm d} \phi = \delta_{ll^\prime} \delta_{mm^\prime}$. From Eq.~\eqref{eq:sm:FDAimp}, we can see that only the mode $(l=0,m=0)$, the $s$-wave mode, is coupled with the closed channel so only these modes need to be considered in the computation.

The Ramsey response may be calculated exactly using a functional determinant approach~\cite{Levitov1996jmp,Levitov1993jetpl,Knap2012prx}. Accordingly, the overlap may be written as
\beq
S(t) = \mathrm{det} \big[1 - \hat{n} + \hat{n} \, e^{i \hat{h}_\rmi t} e^{-i \hat{h} t} \big],
\eeq
where $\hat{n} = 1/ \big[e^{\beta(\hat{h}_\rmi -\mu)} + 1\big]$ is the occupation operator with $\mu$ the chemical potential. $\hat{h}_\rmi$ and $\hat{h}$ are the single-particle counterparts of $\hat{H}_\rmi$ and $\hat{H}$.

\subsection{Relationship between the spectral function and the Ramsey response}

The spectral function ${A}(\omega)$ can be extracted from the Fourier transform of ${S} (t)$ \cite{mahan}:
\begin{align}
{A} (\omega) = {\rm Re} \int_0^\infty \frac{{\rm d} t}{\pi} e^{i\omega t} {S} (t) \simeq \left( \frac{2 \pi}{m_\down T} \right)^{3/2} \sum_{\q n} e^{-\beta \epsilon_{\q \down}} \big|\alpha_{{\bf q};0}^{(n)} \big|^2 \delta (\omega + \epsilon_{\q \down} - E_{\q;n}). \label{eq:sm:Aom_derive}
\end{align}
In the results displayed in the main text, we have convolved the delta function in Eq.~\eqref{eq:sm:Aom_derive} with a Gaussian distribution of width $\sigma$. This mimics an experimental broadening due to the finite length of the radio-frequency pulse. Specifically, in panels (g-i) of Fig.~2 we have applied a broadening of 0.03$/ \tau_{\rm F}$, 0.03$/ \tau_{\rm F}$, and 0.1$/ \tau_{\rm F}$, respectively, which corresponds to the broadening in experiment~\cite{Cetina2016}.
% The first of these is twice what was reported in the experiment~\cite{Cetina2016}, while the other two exactly correspond to the broadening in experiment.

\begin{figure}[hbt]
\centering
\includegraphics[width=1.\textwidth]{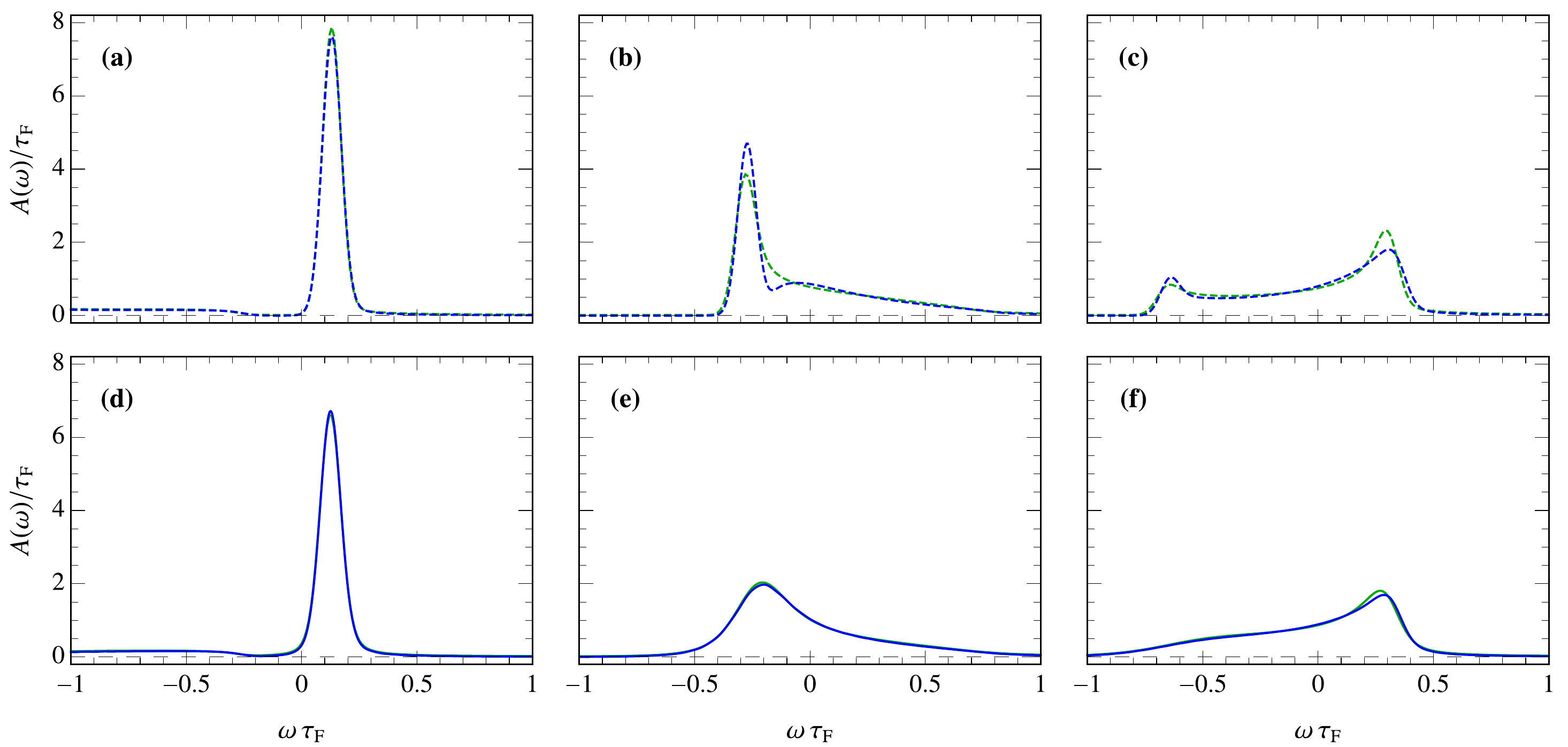}
\caption{The spectral function $A(\omega)$ corresponding to the static impurity Ramsey response shown in Fig.~\ref{fig:benchmarking}. We show our variational results (blue) and exact results (green) for $T=0$ (dashed) and $T=\, 0.2T_{\rm F}$ (solid). From left to right: repulsive, attractive and unitary interactions with the same parameters as in Fig.~\ref{fig:benchmarking}. We have applied a small Gaussian broadening of width $\sigma = 0.04 / \tau_{\rm F}$.}
\label{fig:sm:TBM_spectrum}
\end{figure}

In Fig.~\ref{fig:sm:TBM_spectrum} we show the spectral function for a static impurity calculated within our variational approach using the same interaction parameters as in Fig.~1 of the main text. For the repulsive case, we see that the temperature has little effect on the spectrum; it just slightly decreases the height of the peak. In the attractive case, on the other hand, the distinct attractive peak that exists at negative energy for zero temperature is seen to merge with the continuum at positive energy when the temperature increases. This is due to the temperature causing a broadening of the peak. This is in turn what leads to the significant difference in the phase at zero and finite temperature, see Fig.~\ref{fig:benchmarking}(e). For the unitary case, the clear oscillation of the phase that occurs at zero temperature due to the interference of the repulsive and attractive branches, see Fig.~\ref{fig:benchmarking}(f), is suppressed at finite temperature. This again occurs because the thermal excitations broaden the attractive peak.
%\clearpage
We also compare our results with that of the functional determinant approach (see Fig.~\ref{fig:sm:TBM_spectrum}) and we find good agreement across all temperatures and interaction strengths.

\subsection{Modified Ramsey response in a two-step quench}

In the experiment~\cite{Cetina2016} during the radio-frequency $\pi/2$ pulse that is applied before and after the quench, the system is evolving subject to a Hamiltonian $\hat{H}_1$ with weak interactions for a short period $t_1$. Thus, the corresponding wave function overlap becomes,
\begin{align}
S^\prime_{{\bf q}}(t) = & \, e^{i \epsilon_{{\bf q}\downarrow} (t+2t_1)}\! \big< \hat{c}_{{\bf q}\downarrow} e^{-i\hat{H}_1 t_1} e^{-i \hat{H} t} e^{-i\hat{H}_1 t_1} \hat{c}^\dag_{{\bf q}\downarrow} \big>_{\beta} \nn \\ = & \, e^{i \epsilon_{{\bf q}\downarrow} (t+2t_1)} \big< \big( e^{i \hat{H} t} e^{i\hat{H}_1 t_1} \hat{c}_{{\bf q}\downarrow} e^{-i\hat{H}_1 t_1} e^{-i \hat{H} t} \big) \big(e^{-i\hat{H}_1 t_1} \hat{c}^\dag_{{\bf q}\downarrow} e^{i\hat{H}_1 t_1} \big) \big>_{\beta} \nonumber \\
\equiv &\, e^{i \epsilon_{{\bf q}\downarrow} (t+2t_1)} \avbeta{\hat{c}_{{\bf q}\downarrow,t_1}(t)\, \hat{c}^\dag_{{\bf q}\downarrow,-t_1}(0)}. \label{eq:sm:Sprimet_def}
\end{align}
In going from the first step to the second, we have made use of the fact that $\hat{H}_1$ or $\hat{H}$ is equivalent to $\hat{H}_0$ when acting directly on the Fermi sea. Note that Eq.~\eqref{eq:sm:Sprimet_def} recovers Eq.~\eqref{eq:Sprimet_def}.

Along with Eq.~\eqref{eq:opexp} in the main text, we have
\begin{align}
\hat{\mathbf{c}}_{{\bf q}\down,t_1} & = \sum_n \big<  \hat{c}_{{\bf q}\downarrow} \hat{\phi}_{1,{\bf q}}^{(n)\,\dag} \big>_\beta \hat{\phi}_{1,{\bf q}}^{(n)} \, e^{-i E_{1,{\bf q}}^{(n)} t_1} = \sum_n \alpha_{1,{\bf q};0}^{(n) \, *} \, \hat{\phi}_{1,{\bf q}}^{(n)} \, e^{-i E_{1,{\bf q}}^{(n)} t_1}, 
\end{align}
and then we have
\begin{align}
\hat{\mathbf{c}}_{{\bf q}\down,t_1}(t) & = \sum_l \big< \hat{\bf c}_{{\bf q}\down,t_1} \, \hat{\phi}^\dag_{{\bf q}l}\big>_\beta \hat{\phi}_{{\bf q}l} \, e^{-i E_{\q;l} t} = \sum_{nl} \alpha_{1,{\bf q};0}^{(n) \, *}\big< \hat{\phi}_{1,{\bf q}}^{(n)} \, \hat{\phi}^\dag_{{\bf q}l}\big>_\beta \, \hat{\phi}_{{\bf q}l} \, e^{-i E_{\q;l} t} \, e^{-i E_{1,{\bf q}}^{(n)} t_1}. 
\end{align}
According to Eq.~\eqref{eq:sm:Sprimet_def}, we can easily obtain
\begin{align}
S^{\prime}_{{\bf q}}(t) \simeq &\, e^{i \epsilon_{{\bf q}\downarrow} (t+2t_1)} \sum_{nlm} \alpha_{1,{\bf q};0}^{(n)\,*} \, \alpha_{1,{\bf q};0}^{(m)} \big\langle \hat{\phi}_{1,{\bf q}}^{(n)} \, \hat{\phi}^\dag_{{\bf q}l}\big\rangle_\beta \, \big\langle \hat{\phi}_{{\bf q}l} \,  \hat{\phi}^{m\,\dag}_{1,{\bf q}}\big\rangle_\beta e^{-i \big[ E_{1,{\bf q}}^{(n)}+E_{1,{\bf q}}^{(m)} \big] t_1} e^{-i E_{{\bf q};l} t}. 
\end{align}

\subsection{Effect of temperature on the Ramsey response}

\begin{figure}[hbt]
\centering
\includegraphics[width=\textwidth]{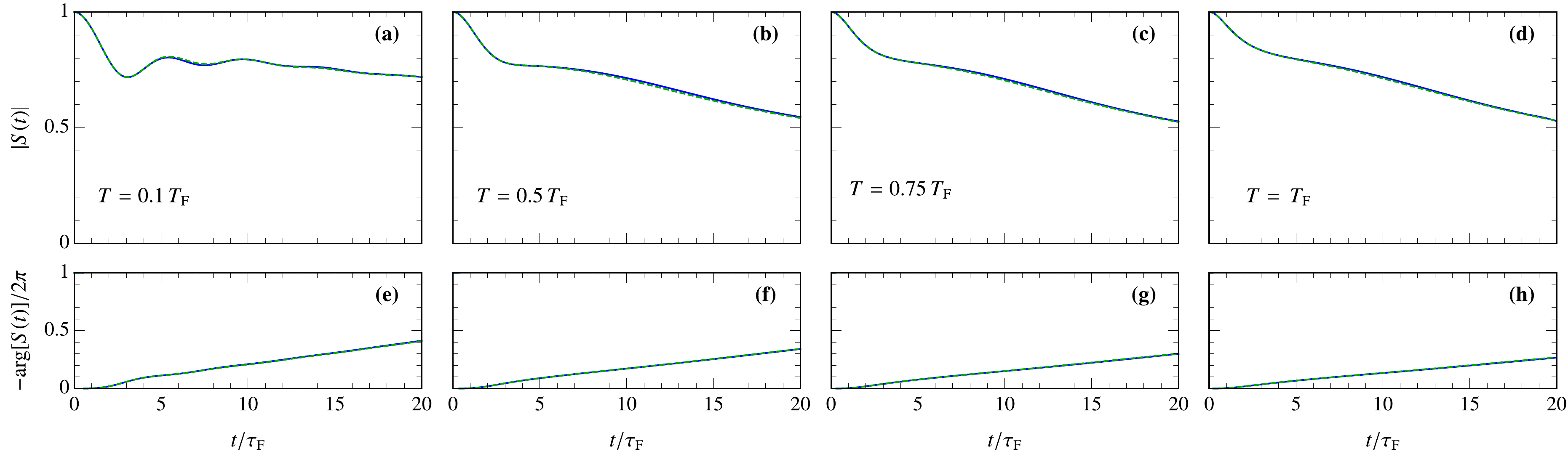}
\caption{Ramsey response for a static impurity for interactions tuned to the repulsive side, with $1/(k_{\rm F}a)=1$ and $k_{\rm F}R^*=1$, which is the also scenario depicted in Fig.~\ref{fig:benchmarking}(a) of the main text. We show the results for different temperatures calculated within the exact approach (green) and the variational method (blue).}
\label{fig:sm:rep}
\end{figure}

\begin{figure}[tbh]
\centering
\includegraphics[width=\textwidth]{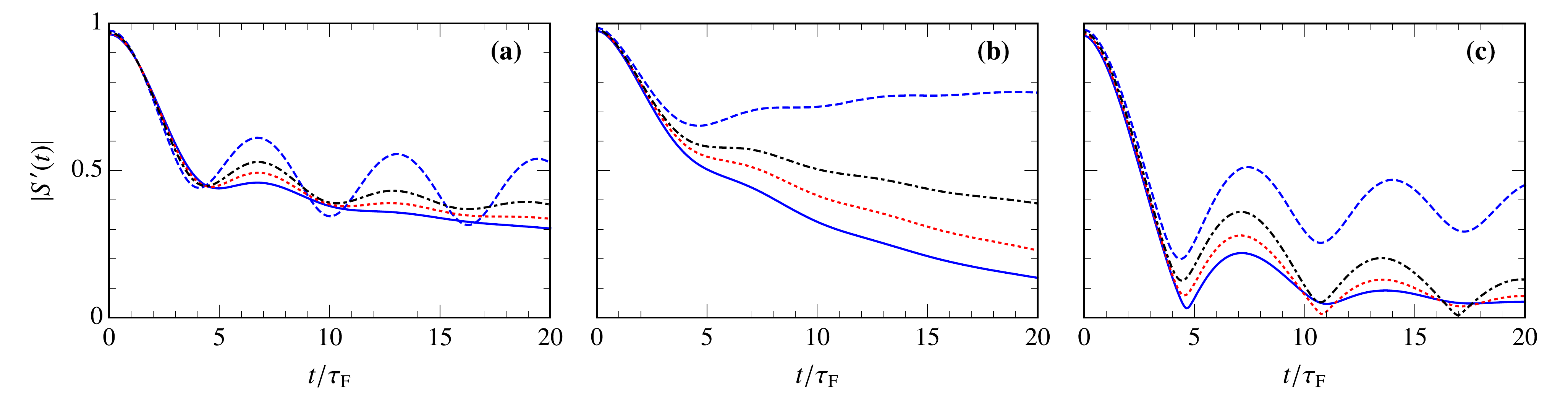}
\caption{Ramsey response for $^{40}$K impurities in a $^6$Li Fermi sea for the same parameters as in Fig.~\ref{fig:ExpFitting} of the main text. From left to right, the experimental parameters are $1/(k_{\mathrm{F}} a) = \{0.23,-0.86,-0.08\},$ $T/T_{\mathrm{F}} = \{0.17,0.16,0.18\},$ and $1/ (k_{\mathrm{F}} a_{1})= \{3.9,-5.8,-4.8\}$. In all panels $k_{\mathrm{F}}R^* = 1.1$ and $t_1 = 4.0 \, \tau_{\mathrm{F}}.$ We show our variational results for four scenarios: A single zero-momentum impurity in a $T=0$ Fermi sea (blue, dashed); a zero-momentum impurity in a finite temperature Fermi sea (red, dotted); a Boltzmann gas of impurities with a $T=0$ Fermi sea (black, dot-dashed); and a Boltzmann gas of impurities in a Fermi sea at finite temperature as shown in the main text (blue, solid).}
\label{fig:sm:tempeffec}
\end{figure}

Figure~\ref{fig:benchmarking} of the main text demonstrates that our approach works very well both at zero temperature and at $T=0.2 \, T_{\rm F}$. However, when comparing with experiment in Fig.~\ref{fig:ExpFitting}(a,g) for the repulsive side, we observe a slight deviation in the decay rate of the repulsive polaron. To further investigate the accuracy of our approach for the repulsive branch, in Fig.~\ref{fig:sm:rep} we compare our variational approach for the infinitely heavy impurity with the exact solution for a range of temperatures.  For this case, we see that our approach essentially exactly captures the dynamics within the time frame shown in the figure.

Finally, in Fig.~\ref{fig:sm:tempeffec}, we see that in order to accurately model the experiment~\cite{Cetina2016}, both the finite-temperature bath and the thermal average over impurity momenta are important. Indeed, taking only one or the other into account yields very comparable results (red dotted, and black dash-dotted lines in Fig.~\ref{fig:sm:tempeffec}), and the final result (blue solid line) is significantly different from both of those approximations.

\end{document}